\documentclass{article}
\usepackage{alife9,epsf,epsfig}
\title{Tierra's missing neutrality: case solved.}
\author{Russell K. Standish\\
School of Mathematics,
University of New South Wales,Sydney, 2052,Australia\\
R.Standish@unsw.edu.au, http://parallel.hpc.unsw.edu.au/rks}

\def\citeyear(#1)#2{(#1)\nocite{#2}}

\begin{document}
\maketitle
\begin{abstract}
  The concept of neutral evolutionary networks being a significant
  factor in evolutionary dynamics was first proposed by Huynen {\em et
    al.} about 7 years ago. In one sense, the principle is easy to
  state --- because most mutations to an organism are deleterious, one
  would expect that neutral mutations that don't affect the phenotype
  will have disproportionately greater representation amongst successor
  organisms than one would expect if each mutation was equally likely.

  So it was with great surprise that I noted neutral mutations being
  very rare in a visualisation of phylogenetic trees generated in {\em
  Tierra}, since I already knew that there was a significant amount of
  neutrality in the Tierra genotype-phenotype map. 

  It turns out that competition for resources between host and
  parasite inhibits neutral evolution.

\end{abstract}

\section{Introduction}

The influence of {\em neutral networks} in evolutionary processes was
first elucidated by Peter Schuster's group in Vienna in
1996 \cite{Huynen-etal96,Reidys-etal96}. Put simply, two {\em
  genotypes} are considered {\em neutrally equivalent} if they map to
the same {\em phenotype}. A {\em neutral network} is a set of
genotypes connected by this neutrality relationship on links with
Hamming distance 1 (i.e. each link of the network corresponds to a
mutation at a single site of the genome). It should be noted that this
definition is subtly different from that employed in Kimura's {\em
  neutral evolution theory} \cite{Kimura83}, as in that theory,
neutrality is defined as equivalence of fitness values, a notion that is
ill-defined in coevolutionary systems. However as phenotypically
equivalent organisms are neutral in Kimura's sense when a fitness
function exists, much of neutral theory can be carried over into
discussion of phenotypic neutrality.

Schuster's group noted that evolution tended to proceed by diffusion
along these neutral networks, punctuated occasionally by rapid changes
to phenotypes as an adaptive feature is discovered. The similarity of
these dynamics with the theory of Punctuated Equilibria
\cite{Eldridge85} was noted by Barnett \citeyear(1998){Barnett98}. It
was also noted that if a {\em giant network} existed that came within
a hop or two of every possible genotype, evolution will be
particularly efficient at discovering solutions, since only a few
non-neutral mutations are needed to reach the optimum solution.

Most work on neutrality in evolution uses the genotype-phenotype
mapping defined by folding of RNA \cite{Schuster-etal94}. This mapping
is implemented in the open source Vienna RNA
package\footnote{http://www.tbi.univie.ac.at/\~{}ivo/RNA}, so is a
convenient and well-known testbed for ideas of neutrality in evolution.

Also in 1996, I developed a definition of the genotype-phenotype
mapping for Tierra, which was first published in 1997
\cite{Standish97b}. I noticed the strong presence of neutrality in
this mapping at that time, which was later exploited to develop a
measure of complexity of the Tierran organism
\cite{Standish99a,Standish03a}. In 2002, I started a programme to
visualise Tierra's phylogenetic trees and neutral networks
\cite{Standish02c} in order to ``discover the unexpected''.  Two key
findings came out of this: the first being that Tierra's
genebanker\footnote{The {\em genebanker} is a database in which Tierra
  stores the genotypes that arise during evolution.} data did not
provide clean phylogenetic trees, but had loops, and consisted of many
discontinuous pieces. This later turned out to be due to Tierra's
habit of reusing genotype labels if those genotypes were not saved in
the genebanker database. This might happen if the population count of
that genotype failed to cross a threshold.  This is all very well,
except that a reference to that genotype exists in the parent field of
successor genotypes. The second big surprise was the paucity of
neutral mutations in the phylogenetic tree. We expect most mutations
to an organism to be deleterious, and so expect that neutral mutations
will have disproportionately greater representation amongst successor
genotypes than one would expect if each mutation was equally likely.

\section{Neutrality in Tierra}

{\em Tierra} \cite{Ray91} is a well known artificial life system in which
small self-replicating computer programs are executed in a specially
constructed simulator. These computer programs (called digital
organisms, or sometimes ``critters'') undergo mutation, and radically
novel behaviour is discovered, such as {\em parasitism} and {\em
  hyperparasitism}.

It is clear what the genotype is in Tierra, it is just the listing of
the program code of the organism. The phenotype is a more diffuse
thing, however. It is the resultant effect of running the computer
program, in all possible environments. Christoph Adami defined this
notion of phenotype for a similar artificial life system called {\em
  Avida} \cite{Adami98a}. In Avida, things are particularly simple, in
that organisms either reproduce themselves at a fixed replication rate,
or don't as the case may be, and optionally perform range of
arithmetic operations on special registers (defined by the experimenter).

In Tierra, organisms do interact with each other via a template
matching mechanism. For example, with a branching instruction like
\verb+jmpo+, if there is a sequence of \verb+nop0+ and \verb+nop1+
instructions (which are no-operations) following the branch, this
sequence of 1s and 0s is used as a template for determining where to
branch to. In this case the CPU will search outwards through memory
for a complementary sequence of \verb+nop0+s and \verb+nop1+s. If the
nearest complementary sequence happens to lie in the code of a different
organism, the organisms interact.

To precisely determine the phenotype of a Tierran organism, one would
need to execute the soup containing the organism and all possible
combinations of other genotypes. Whilst this is a finite task, it is
clearly astronomically difficult. One means of approximation is to
consider just interaction of pairs of genotypes (called a tournament).
Most Tierran organisms interact pairwise --- very few triple or higher
order interactions exist. Similarly, rather than running tournaments
with all possible genotypes, we can approximate matters by using the
genotypes stored in a genebanker database after a Tierra
run. In practice, it turns out that various measures, such as the
number of neutral neighbours, or the total complexity of an organism
are fairly robust with respect to the exact set of organism used for
the tournaments.

So the procedure is to pit pairwise all organisms in the genebanker
against themselves, and record the outcome in a table (there is a
small number of possible outcomes, which is detailed in
 \cite{Standish97b}). A row of this table is a phenotypic signature for
the genotype labeling that row. We can then eliminate those genotypes
with identical signatures in favour of one canonical genotype. This
list of unique phenotypes can be used to define pragmatically a test
for neutrality of two different genotypes, that may have generated by
mutation from genotypes recorded in the genebanker. Pit each organism against
the list of unique phenotypes, and if the signatures match, we have
neutrality.  The source code for this experiment is available from the
author's
website.\footnote{http://parallel.hpc.unsw.edu.au/getaegisdist.cgi/getsource/eco-tierra.3,
  version 3.D3}

Tierra has three different modes of mutation:
\begin{description}
\item[Cosmic Ray] A site of the soup is randomly chosen and mutated;
\item[Copy] Data is mutated during the copy operation;
\item[Flaw] Instructions occasionally produce erroneous results
\end{description}
Furthermore, in the case of cosmic ray and copy mutations, a certain
proportion of mutations involve bit flips, rather than opcodes being
substituted uniformly. This proportion is set as a parameter in the
soup\_in file (\verb+MutBitProp+) --- in these experiments, this
parameter is set to zero.

In order to study the issue of whether neutrality is greater or less
than expected in Tierra, I generated three datasets with each of the 3
modes of mutation operating in isolation. The sizes of each data set
was 69,139, 87,003 and 198,982 genotypes respectively, 
generated over a time period of about 1000 million executed instructions.
Genebanker's threshold was set to zero, so all genotypes were
captured. This led to a proper phylogenetic tree. After performing a
neutrality analysis, a set of 83, 86 and 158 unique phenotypes was extracted as
the test set for the tournaments.

\begin{figure}
\epsfxsize=.5\textwidth\epsfbox{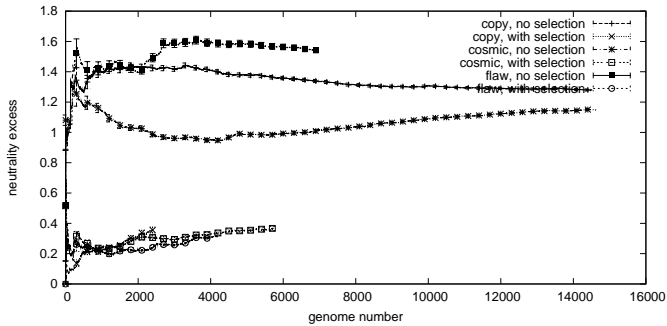}
\caption{Running average of neutrality excess ($\langle
  r_i\rangle$). Genomes are ordered according to size, and neutrality
  excess is averaged over all genomes to the left of that data
  point. Three different datasets are analysed, with each of the three
  modes of mutation turned on. Then the datasets are further filtered
  to only include offspring whose maximum population count is greater
  than 1, i.e. selection is operating.
}
\label{nn-ratio}
\end{figure}

Since the neighbourhood size increases exponentially with
neighbourhood diameter, I restrict analysis to single site, or point mutations.
In each data set, around 7\% of these genotypes were
created  by a mutation at a single
site and were neutrally equivalent to its parent. 
 For each of these, I compute the number of neutral
neighbours $n_i$ existing in the 1 hop neighbourhood of the parent
genotype $i$. The 1 hop neighbourhood size is $32^{\ell_i}$, where $\ell_i$ is
the length of the genome. For a given parent $i$, the ratio 
\begin{equation}\label{neutrality excess}
r_i=\frac{\nu_i32^{\ell_i}}{o_in_i}
\end{equation}
gives the proportion of neutral links actually followed relative to
the number of neutral links available ({\em neutrality excess}), where
$\nu_i$ is the number of neutrally equivalent offspring, and $o_i$ the
total number of offspring and $n_i$ the size of the 1 hop neutral
neighbourhood. Fig.~\ref{nn-ratio} shows the running average of this
quantity over these transitions, with the genotypes numbered in size order.

Since all daughter genotypes are recorded, no selection is operating.
In this case, one would expect that the proportion of neutral variants
seen should be identical to the proportion of neutral variants within
the 1 hop neighbourhood, and hencethe neutrality excess should be
identical to 1. However, in the case of instruction and cosmic ray
flaws, not every daughter genome will make it into the genebanker. In
the case of instruction flaws, it is rather unpredictable what the
effect is. In the case of cosmic ray mutations, 50\% of time one would
expect the parent to be mutated, rather than the daughter. In the case
of a mutation affecting a crucial gene of a parent genotype, the
organism may not be able to reproduce at all, thus favouring neutral
mutations. Only copy mutations should affect all sites of the genome
equally, leading to a neutrality excess equal to one. The measured
value, however is about 1.3, substantially greater than one. The
reason for this is not known at this point in time.

The datasets were further subsetted to include just those
transitions whose daughter organism successfully reproduced, i.e. with a
maximum population count greater than 1. The neutrality excess in this
case is substantially less than 1, so something in Tierran evolution
is favouring non-neutral evolution.

\section{Competition Effects}

Consider a single species ecosystem with logistic dynamics:
\begin{equation}
\dot{x} = rx(1 - x/K),
\end{equation}  
where $x$ is the population size, $r$ the net reproductive rate and
$K$ the carrying capacity. A phenotypically equivalent genotype
attempting to invade this ecosystem will have the following dynamics:
\begin{equation}
\dot{x'} = rx'(1 - x/K) \approx 0,
\end{equation}
($x'$ being the population size of the invading genotype) as $x\approx
K$ at equilibrium. So there is a substantial likelihood that the
neutral variant fails to invade the ecosystem.

This argument is of course an extreme case. Stochastic effects due to
finite population sizes will increase the chances of a neutral variant
invading the ecosystem, however the point still remains that the
neutral variant is not on an equal footing as the incumbent.

In Tierra, however, there is an age structure in the population, with
organisms being placed in a {\em reaper queue}, from which the oldest
organisms are selected when death is required. This fact alone implies
that neutral variants of self-replicating organisms will successfully
replicate, and hence cannot be responsible for the neutrality
deficiency.

However, consider a Tierran ecosystem consisting of hosts and
parasites, where the parasite require the presence of a host organism
within a certain distance of itself in the soup, in order for the
parasite to replicate. Since parasitic organisms replicate faster than
the hosts (due to their smaller program lengths), they tend to
displace host organisms until there are not enough hosts to go
around. At which point, the parasite's fecundity drops. At
equilibrium, the effective reproductive rates of host and parasite are
equal.

A neutral variant will therefore be quite likely to not have a
suitable host in its neighbourhood to allow it to
replicate. Consequently, neutral evolution is suppressed amongst
parasites. In the next section I will test this idea by setting up an
artificial host-parasite coevolutionary system, using the well known
RNA genotype-phenotype map.

\section{Vienna RNA Folding Experiments}

It is quite well known that evolution using the RNA folding map
\cite{Schuster-etal94} exhibits a great deal of neutrality, at least
for a standard genetic algorithm optimising a well defined fitness
function. Until now, evolutionary systems based on the RNA map exhibit
the unsurprising result of neutrality excess defined by eq
(\ref{neutrality excess}) being greater than or equal to 1. I now
present results of an RNA map experiment that demonstrates neutrality
supression ($r_i<1$), based on the resource competition explanation
elabortaed earlier.  We need two types of organism (host and parasite)
competing for a fixed space that can support $N=100$ organisms.
Parasites can only reproduce if they are situated next to a host
(neighbourhood size $\nu=2$), but reproduce twice as fast as the host
type.

Once an organism has reproduced, it replaces the least fit
organism. Fitness is determined by how close the parasitic phenotype
is to any hosts in the neighbourhood of the parasites, and decreases
in a similar way with the similarity of the parasites in the
neighbourhood for host organisms:
\begin{eqnarray}
F_\mathrm{h} &=& 1-\frac1{\nu\ell} \sum_{i\in {\cal P}_\nu}d(i,h)\\
F_\mathrm{p} &=& \frac\rho{\nu\ell}\sum_{i\in {\cal H}_\nu}d(i,p)
\end{eqnarray}
where $h$ and $p$ are host and parasite genotypes respectively, ${\cal
  H}_\nu$ and ${\cal P}_\nu$ the set of hosts and parasites
respectively within the neighbourhood of size $\nu$ of $p$ and $h$
respectively. $d(i,j)$ is the string edit distance between the
phenotypes\footnote{The {\em string edit} distance is related to the
  Hamming distance (no. of base pairs that differ between two
  strings), but allows for gaps in the strings. Given a set of edit
  operations (eg insertionsq and deletions) and edit costs, the edit
  distance is given by the minimum sum of the costs along an edit path
  converting one object into the other. Please consult the Vienna RNA
  package documentation for a precise definition of string edit
  distance}, and $\ell$ is the gene length (set equal to 20 for all
organisms in this experiment).

The factor $\rho$ adjusts the relative dominance of parasites over
hosts. Set it too low, and hosts will eliminate the parasites by virtue
of replacing then when replicating. Set it too high, and hosts will
only be competing with themselves. In this experiment, a value of $\rho=3.1$
was found to give intermediate behaviour.

An alternative version of this experiment where organisms were
selected at random for death, rather than according to a fitness
relationship showed similar dynamics, although the neighbourhood size
$\nu$ needed to be increased to 4 to allow a stable population of
parasites to persist.

\begin{figure}
\epsfxsize=.5\textwidth\epsfbox{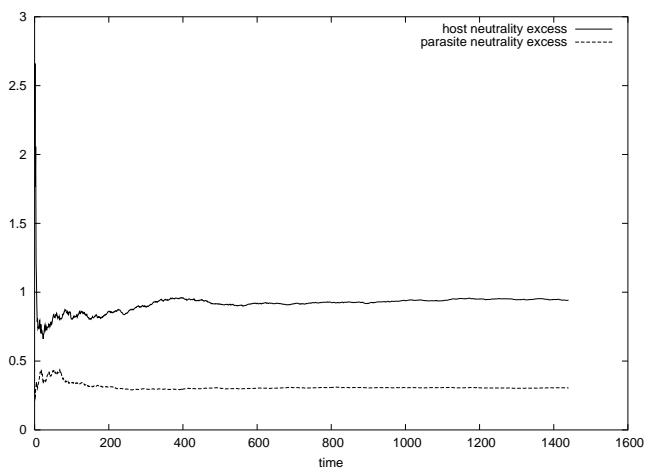}
\caption{Neutrality excesses for RNA folding host-parasite system as
  described in the text.}
\label{rnafold}
\end{figure}

Figure \ref{rnafold} shows the neutrality excess for this
experiment. The model will consistently produce a neutrality
deficiency for the parasites over a broad range of model
parameters. If the parameter $\rho$ is set too high, the hosts will
compete strongly with themselves, suppressing neutrality in the host
population also.

Source code for this experiment is available from the author's
website.
\footnote{http://parallel.hpc.unsw.edu.au/getaegisdist.cgi/getsource/rnafold/, 
  version D1}

\section{Conclusion}

The suppression of neutrality in Tierran evolution is a real
effect. An explanation couched in terms of host parasite competition
was found, and a model was constructed using the well-known RNA folding
map that illustrated this explanation.

This finding is potentially important. It has been argued that neutral
diffusion is an important feature of evolutionary processes allowing
efficient search of phenotype space. The sort of competition effects
seen here to impede neutral diffusion are characteristic of climax
ecosystems. This would imply that disturbed ecosystems will have
greater evolvability than climax systems. This ``brake'' on neutral
diffusion being released during times of environmental stress could
provide an alternative explanation for the patterns of adaptive
radiation seen after mass extinction events.

\section*{Acknowledgments}

I would like to thank the {\em Australian Centre for Advanced
  Computing and Communications} for a grant of computing time used in
  this project.

\bibliographystyle{alife9}
\bibliography{rus}

\begin{thebibliography}{}

\bibitem[Adami, 1998]{Adami98a}
Adami, C. (1998).
\newblock {\em Introduction to Artificial Life}.
\newblock Springer.

\bibitem[Barnett, 1998]{Barnett98}
Barnett, L. (1998).
\newblock Ruggedness and neutrality --- the {$NKp$} family of fitness
  landscapes.
\newblock In Adami, C., Belew, R., Kitano, H., and Taylor, C., editors, {\em
  Artificial Life {VI}}, pages 18--27, Cambridge, Mass. MIT Press.

\bibitem[Eldridge, 1985]{Eldridge85}
Eldridge, N. (1985).
\newblock {\em Time Frames --- The Rethinking of Darwinian Evolution and the
  Theory of Punctuated Equilibria}.
\newblock Simon and Schuster, New York.

\bibitem[Huynen et~al., 1996]{Huynen-etal96}
Huynen, M., Stadler, P.~F., and Fontana, W. (1996).
\newblock Smoothness within ruggedness: The role of neutrality in adaption.
\newblock {\em Proc. Nat. Acad. Sci. USA}, 93:397.

\bibitem[Kimura, 1983]{Kimura83}
Kimura, M. (1983).
\newblock {\em The Neutral Theory of Molecular Evolution}.
\newblock Cambridge UP, Cambridge.

\bibitem[Ray, 1991]{Ray91}
Ray, T. (1991).
\newblock An approach to the synthesis of life.
\newblock In Langton, C.~G., Taylor, C., Farmer, J.~D., and Rasmussen, S.,
  editors, {\em Artificial Life {II}}, page 371. Addison-Wesley, Reading, Mass.

\bibitem[Reidys et~al., 1997]{Reidys-etal96}
Reidys, C., Kopp, S., and Schuster, P. (1997).
\newblock Evolutionary optimization of biopolymers and sequence structure maps.
\newblock In Langton, C. and Shimohara, K., editors, {\em Artificial Life {V}},
  page 379, Cambridge, Mass. MIT Press.

\bibitem[Schuster et~al., 1994]{Schuster-etal94}
Schuster, P., Fontana, W., Stadler, P.~F., and Hofacker, I.~L. (1994).
\newblock From sequences to shapes and back: A case study in {RNA} secondary
  structures.
\newblock {\em Proc. Royal Soc. London B}, 255:279--284.

\bibitem[Standish, 1997]{Standish97b}
Standish, R.~K. (1997).
\newblock Embryology in {T}ierra: A study of a genotype to phenotype map.
\newblock {\em Complexity International}, 4.

\bibitem[Standish, 1999]{Standish99a}
Standish, R.~K. (1999).
\newblock Some techniques for the measurement of complexity in {Tierra}.
\newblock In Floreano, D., Nicoud, J.-D., and Mondada, F., editors, {\em
  Advances in Artificial Life: 5th European Conference, ECAL 99}, volume 1674
  of {\em Lecture Notes in Computer Science}, page 104, Berlin. Springer.

\bibitem[Standish, 2003]{Standish03a}
Standish, R.~K. (2003).
\newblock Open-ended artificial evolution.
\newblock {\em International Journal of Computational Intelligence and
  Applications}, 3:167.

\bibitem[Standish and Galloway, 2002]{Standish02c}
Standish, R.~K. and Galloway, J. (2002).
\newblock Visualising {Tierra's} tree of life using {Netmap}.
\newblock In Bilotta, E. et~al., editors, {\em ALife VIII Workshop
  proceedings}, page 171.
\newblock http://alife8.alife.org.

\end{thebibliography}

\end{document}